\documentclass[12pt,a4paper]{article}
\usepackage{amsmath,amssymb,bm,ascmac,bbm,mathtools}
\usepackage[dvipdfmx, usenames]{color}
\usepackage[dvipdfmx]{graphicx}
\usepackage{xcolor}
\usepackage{here}
\usepackage{authblk}
\usepackage{hyperref}
\usepackage[hang,small,bf]{caption}
\usepackage[subrefformat=parens]{subcaption}
\usepackage{dcolumn}

\newcommand{\tr}{\text{tr}}

\newcommand{\mcal}{\mathcal}
\newcommand{\mbb}{\mathbb}
\newcommand{\mfrak}{\mathfrak}

\newcommand{\rank}{\text{rank}}

\newcommand{\fp}{\text{FPdim}}
\newcommand{\fib}{\text{Fib}}

\newcommand{\vect}{\text{Vec}}

\newcommand{\Hom}{\underline{\text{Hom}}}

\begin{document}
\title{Ground state degeneracy and module category}
\author{Ken KIKUCHI} 
\affil{Department of Physics, National Taiwan University, Taipei 10617, Taiwan}
\date{}
\maketitle

\begin{abstract}
We develop a systematic method to classify connected étale algebras $A$'s in (possibly degenerate) pre-modular category $\mathcal B$. In particular, we find the category of $A$-modules, $\mathcal B_A$, have ranks bounded from above by $\lfloor\text{FPdim}(\mathcal B)\rfloor$. For demonstration, we classify connected étale algebras in some $\mathcal B$'s, which appear in physics. Physically, the results constrain (or fix) ground state degeneracies of (certain) $\mathcal B$-symmetric gapped phases. We study massive deformations of rational conformal field theories such as minimal models and Wess-Zumino-Witten models. In most of our examples, the classification suggests the symmetries $\mathcal B$'s are spontaneously broken.
\end{abstract}

\makeatletter
\renewcommand{\theequation}
{\arabic{section}.\arabic{equation}}
\@addtoreset{equation}{section}
\makeatother

\section{Introduction}
The goal of this paper is to classify connected étale algebras in (possibly degenerate) pre-modular fusion category. This problem is actively studied in mathematics \cite{DMNO10,DMO11}, while our motivation comes from physics, quantum field theory (QFT). One of main goals in QFT is to figure out long distance (or infrared, IR) behaviors of a theory defined at a short distance (or ultraviolet, UV).\footnote{Keeping track of the scale dependence is called renormalization group (RG) flow.} The IR behaviors are basically distinguished by two criteria: gap and spontaneous symmetry breaking (SSB). IR theories are either gapped or gapless, and surviving UV symmetries are either preserved or spontaneously broken. Therefore, our central goal in QFT is to show in which quadrant a given UV symmetry belongs.

For example, quantum chromodynamics (QCD) also fits in this structure. The theory is defined at UV as $SU(3)$ gauge theory with quarks, and we have been trying to prove its IR behaviors. The IR theory is believed to be gapped, confined (for, say, adjoint quarks), and its chiral symmetry is spontaneously broken (when quarks are massless). The confinement is defined as preserved one-form symmetry. Hence, the conjecture says the one-form symmetry belongs to the quadrant $(\text{gapped,preserved})$, and chiral symmetry belongs to $(\text{gapped,spontaneously broken})$. Proving this IR behavior is still an open problem.

In order to explore the IR behaviors, symmetries have been playing central roles. In the modern definition \cite{GKSW14}, symmetries are generated by topological operators supported on submanifolds with some codimensions. The defining topological nature should make it clear why they constrain IR dynamics; being topological, they do not depend on distances, and surviving UV symmetries persist in IR theories. In particular, 't Hooft anomalies \cite{tH80} associated to symmetries have been used extensively to constrain IR behaviors.

In general, such generalized symmetries are described by monoidal categories, and the anomalies are encoded in their associativity structures \cite{BT17}. However, categorical symmetries in physical problems typically have more structures, such as braiding \cite{MS1,MS2}. It was found that these additional structures can constrain IR behaviors stronger than anomaly alone \cite{KK21,KKSUSY,KK22II,KK22free,NK22,KKWZW}. These studies focused on cases when the IR theories are gapless.

In this paper, we focus on gapped cases just like QCD. With this assumption, our task to figure out IR behaviors is simplified; we just ask whether surviving UV symmetries are spontaneously broken or not. In order to answer this question, we study one of the fundamental observables in gapped phases, ground state degeneracies (GSDs). If one finds GSD is larger than one, then it signals SSB.\footnote{If the gapped theory is a `direct sum' of more than one theories, one can have $\text{GSD}>1$ without SSB.\label{directsumtheory}} Therefore, the goal of this paper is to constrain (or fix) GSDs in gapped phases.\footnote{Previously, GSDs have been constrained employing the Lieb-Schultz-Mattis theorem \cite{LSM61} and its generalizations \cite{AL86,Osh00,H03,FO15,YHO18}. They have used anomalies. Also see \cite{TW19,HLS21}. However, algebras are known as ways to gauge \cite{FRS02,CR12,BT17}, and they are by definition anomaly-free. Therefore, our constraints are philosophically orthogonal to these studies.}

In two dimensions, a nice one-to-one correspondence is known. Two-dimensional gapped phases with $\mcal C$ symmetry stand in bijection with $\mcal C$-module categories \cite{TW19,HLS21}
\begin{equation}
    \{\text{2d }\mcal C\text{-symmetric gapped phases}\}\cong\{\mcal C\text{-module categories }\mcal M\}.\label{1to1}
\end{equation}
(Some definitions are reviewed in section \ref{pre}.) In particular, GSD in the LHS is given by rank\footnote{The number of (isomorphism classes of) simple objects in $\mcal M$ is called rank and denoted $\rank(\mcal M)$.} of $\mcal M$, $\rank(\mcal M)$, in the RHS. In this way, we can translate the physical problem in the LHS to a mathematical problem in the RHS. The classification of $\mcal C$-module categories is a well-studied mathematical problem initiated by Ostrik \cite{O01}. Once we have classified module categories, we immediately learn what are possible GSDs:
\begin{equation}
    \text{GSD}\in\{\rank(\mcal M)|\mcal C\text{-module categories }\mcal M\}.\label{possibleGSD}
\end{equation}
For instance, if a symmetry category $\mcal C$ only has module categories with ranks two and four, we immediately find GSDs should be two or four. (See, say, our fourth example.) Just as in this example, if there is no rank one $\mcal C$-module category, the classification result mathematically shows $\mcal C$ should be spontaneously broken in gapped phases (assuming $\mcal M$'s are indecomposable).

Note the universal nature of this method. Once we have classified $\mcal C$-module categories, the results apply not only to a physical system we are studying, but also to other systems with the same\footnote{In particular, as we will see, braidings have to be the same.} symmetry category $\mcal C$. In the example above, one system may realize $\text{GSD}=2$, while another system may realize $\text{GSD}=4$.

In short, if we focus on two-dimensional $\mcal C$-symmetric gapped phases, the physical problem on SSB can be answered by classifying $\mcal C$-module categories. However, classification of generic $\mcal C$-module categories is still difficult. Thus, we take an indirect path; we classify connected étale algebras $A$'s in pre-modular category $\mcal C$. (For classifications of connected étale algebras in minimal models and Wess-Zumino-Witten (WZW) models, see \cite{KL02,G23}.\footnote{We thank Victor Ostrik for teaching these papers to us.}) How are the two classification problems related? If we assume module categories $\mcal M$'s are finite, then it is known that there exists an algebra $A\in\mcal C$ such that the category of (right) $A$-modules $\mcal C_A$ is equivalent to $\mcal M$. Therefore, in this paper, expecting we could relax some assumptions in the future, we attempt to classify connected étale\footnote{When an ambient category $\mcal B$ is modular, $A\in\mcal B$ can be interpreted as condensing object, and physically natural conditions demand $A$ be connected étale \cite{K13}.} algebras $A$'s, and translate the results to physics side.

Before we explain our method, we have to make a few comments on our setup. First, we assume our symmetry categories are pre-modular. Since they are equipped with braidings, we write them $\mcal B$ instead of $\mcal C$. Our assumption gives us a lot with little loss of generality (in physics); the pre-modular symmetry categories are common in two dimensions. We usually pick rational conformal field theories (RCFTs) as UV theories. Then, their deformations preserve pre-modular fusion categories $\mcal B$'s. Thanks to their braidings, we can discuss commutativity of $A\in\mcal B$. Since $\mcal B$ is finite, a category of (right) $A$-modules $\mcal B_A$ is also finite, and it is guaranteed to be a fusion category (for separable algebras). Thanks to the rank-finiteness of fusion categories and our upper bound on their ranks, we are left with only finitely many candidates for $\mcal B_A$. This fact makes our classification problem manageable. Second, since we are interested in SSB of $\mcal B$, we assume $\mcal M$'s are indecomposable. Then, $\text{GSD}>1$ implies SSB. (See the lemma in section \ref{pre}.) When we discuss physical implications of results, we further assume $\mcal M$ be finite. Then the equivalence $\mcal M\simeq\mcal B_A$ for some $A\in\mcal B$ connects two classification problems of $\mcal B$-module categories and connected étale algebras. Note, however, we do \textit{not} assume $\mcal B$ be pseudo-unitary. Thus, our method also works for non-unitary RCFTs. We demonstrate our classification procedures in massive deformations of minimal models and WZW models with or without unitarity. We constrain (or fix) GSDs in the IR, and also discuss SSB of surviving UV symmetries.

\section{Classification}
In this section, we first review some facts necessary to understand our method. Based on the knowledge, we explain our classification procedure in the second subsection. In the third subsection, we study physically motivated examples.

\subsection{Preliminary}\label{pre}
Here, we give a minimal review. For more details, see standard textbooks \cite{ML98,R16,EGNO15}. (For a one-page introduction to category theory, see the appendix A of \cite{KKARG}.)

A monoidal category $\mcal C$ is equipped with a bifunctor $\otimes:\mcal C\times\mcal C\to\mcal C$ called monoidal product. (Physicists usually call it fusion product.) It is subject to two coherence conditions, pentagon and unit axioms. An existence of a unit object $1\in\mcal C$ obeying $\forall c\in\mcal C,\ 1\otimes c\cong c\cong c\otimes 1$ is part of the axioms. In a monoidal category $\mcal C$, an object $c^*\in\mcal C$ is called a left dual of $c\in\mcal C$ if there exist evaluation $c^*\otimes c\to1$ and coevaluation $1\to c\otimes c^*$ morphisms subject to some axioms. Similarly, one defines right duals. An object $c\in\mcal C$ with left dual is called self-dual if $c\cong c^*$. (Similarly for right duals.) A category is called rigid if all objects have both left and right duals. A rigid monoidal category is called pivotal if it is equipped with a monoidal natural isomorphism $a:id_{\mcal C}\cong(-)^{**}$ called pivotal structure. With a pivotal structure $a$, one defines left and right quantum (or categorical) traces $\tr^{L/R}(a):1\to1$. A pivotal structure is called spherical if $\forall c\in\mcal C$, $\tr^L(a_c)=\tr^L(a_{c^*})$. Since $\tr^L(a_c)=\tr^R(a_{c^*})$, left and right quantum traces coincide in spherical categories. Thus, we simply write them $\tr(a)$.

There is another important bifunctor. It is usually denoted additively, $\oplus:\mcal C\times\mcal C\to\mcal C$, and called direct sum. A category with direct sum obeying some conditions is called additive. An additive category $\mcal C$ is called $\Bbbk$-linear if $\forall c,c'\in\mcal C$, a collection $\mcal C(c,c')$ of morphisms from $c\in\mcal C$ to $c'\in\mcal C$ is a $\Bbbk$-vector space. Thus, a $\Bbbk$-linear category has a zero object $0\in\mcal C$. A nonzero object $c\in\mcal C$ is called simple if only $0$ and $c$ are its subobjects. The number of isomorphism classes of simple objects in a category $\mcal C$ is called rank and denoted $\rank(\mcal C)$. A category is called semisimple if any objects can be decomposed into direct sums of finitely many simple objects. Mathematically, the direct sum is defined as (co)limit. Thus, a direct sum $c\cong c_1\oplus c_2\oplus\cdots\oplus c_n$ is equipped with product projections $p_i:c\to c_i$ and coproduct injections $\iota_i:c_i\to c$. They obey
\begin{equation}
    p_i\cdot\iota_j\cong\delta_{i,j}id_{c_i},\quad\sum_{i=1}^n\iota_i\cdot p_i\cong id_c.\label{pirelations}
\end{equation}
An additive category with canonical decompositions is called abelian.

A $\Bbbk$-linear abeilan category $\mcal C$ is called locally finite (or artinian) if the following two conditions are satisfied: i) $\forall c,c'\in\mcal C$, $\mcal C(c,c')$ is a finite dimensional $\Bbbk$-vector space, and ii) $\forall c\in\mcal C$ has finite length. A locally finite $\Bbbk$-linear abelian category $\mcal C$ is called finite if two additional conditions are met: i) $\mcal C$ has enough projective objects, and ii) $\rank(\mcal C)$ is finite. For a $\Bbbk$-linear abelian category with finite length, one can define Grothendieck group $\text{Gr}_0(\mcal C)$ as the free abelian group generated by isomorphism classes of simple objects. If $\mcal C$ further has monoidal product, decomposition of $c\otimes c'$ into simple objects introduces $\text{Gr}_0(\mcal C)$ a product to make it an $\mbb N$-ring. The ring is called the Grothendieck ring $\text{Gr}(\mcal C)$. Here is one important fact. Since it is an $\mbb N$-ring, any objects $c$ in $\mcal C$ with a Grothendieck ring are decomposed into simple objects with $\mbb N$ coefficients. In other words, a category $\mcal C$ with Grothendieck ring $\text{Gr}(\mcal C)$ admits non-negative integer matrix representation (NIM-rep). The matrix is defined as
\begin{equation}
    (N_i)_{jk}:={N_{ij}}^k,\label{Nijk}
\end{equation}
where ${N_{ij}}^k\in\mbb N$ is the coefficient
\begin{equation}
    c_i\otimes c_j\cong\bigoplus_k{N_{ij}}^kc_k.\label{fusionrule}
\end{equation}
NIM-reps have been studied actively when $\mcal C$ is modular (see below for the definition), but what we are saying is that it can be defined more generally. Especially, we will use the fact that actions of pre-modular categories $\mcal B$'s should form NIM-reps. Since their matrix elements are non-negative, we can apply the Perron-Frobenius theorem to obtain the largest positive eigenvalue. It is called the Frobenius-Perron dimension, and denoted $\fp_{\mcal C}(N_i)$ or $\fp_{\mcal C}(c_i)$. We added the subscript because monoidal products depend on in which ambient category $\mcal C$ one is working. Abstractly, the Frobenius-Perron dimension is a ring homomorphism
\begin{equation}
    \fp:\text{Gr}(\mcal C)\to\mbb C.\label{FPhom}
\end{equation}
Therefore, the Frobenius-Perron dimension of a direct sum $c\cong c_1\oplus\cdots\oplus c_n$ is given by
\begin{equation}
    \fp_{\mcal C}(c)=\sum_{i=1}^n\fp_{\mcal C}(c_i).\label{FPdirectsum}
\end{equation}
These are Frobenius-Perron dimensions of objects. Additionally, we define that of the category itself by
\begin{equation}
    \fp(\mcal C):=\sum_{i=1}^{\rank(\mcal C)}\left(\fp_{\mcal C}(c_i)\right)^2.\label{FPcategory}
\end{equation}
One essential fact we are going to use is \cite{ENO02,EGNO15}
\begin{equation}
    \forall c\in\mcal C,\quad\fp_{\mcal C}(c)\ge1.\label{FPge1}
\end{equation}

Let $\mcal C$ be a locally finite $\Bbbk$-linear abelian rigid monoidal category. Namely, $\mcal C$ is equipped with three operations, scalar multiplication over $\Bbbk$, direct sum $\oplus$, and monoidal product $\otimes$. The category $\mcal C$ is called a multitensor category if the monoidal product $\otimes$ is bilinear. A multitensor category with simple identity object $1\in\mcal C$ is called tensor category. A multifusion category is a finite semisimple multitensor category. A multifusion category with simple identity object is called fusion category (FC).
\begin{table}[H]
\begin{center}
\begin{tabular}{c|c|c}
    Simple $1\backslash$Finite semisimple&No&Yes\\\hline
    No&multitensor&multifusion\\\hline
    Yes&tensor&fusion
\end{tabular}.
\end{center}\label{multitensor}
\caption{Names of locally finite $\Bbbk$-linear abelian rigid monoidal category with bilinear $\otimes$}
\end{table}
\hspace{-17pt}In this paper, we focus on FCs. An FC over $\mbb C$ is called pseudo-unitary if the categorical (or global) dimension defined by $D^2(\mcal C):=\sum_{i=1}^{\rank(\mcal C)}\tr^L(a_{c_i})\tr^L(a_{c^*_i})$ equals the Frobenius-Perron dimension, $D^2(\mcal C)=\fp(\mcal C)$.

A braiding $c$ in a monoidal category $\mcal B$ is a natural isomorphism $c_{b_i,b_j}:b_i\otimes b_j\cong b_j\otimes b_i$ subject to hexagon axioms. (In order to avoid confusion, we write braided categories and their simple objects as $\mcal B$ and $b_i,b_j,\dots$, respectively.) A fusion category with braiding is called braided fusion category (BFC). A BFC is called pre-modular if it is spherical. This is our ambient category. Let $\mcal B$ be a BFC. Two objects $b_i,b_j\in\mcal B$ are said to commute if $c_{b_j,b_i}\cdot c_{b_i,b_j}\cong id_{b_i\otimes b_j}$. The collection of objects commuting with all $b\in\mcal B$ is called the symmetric center
\begin{equation}
    Z_2(\mcal B):=\{b\in\mcal B|\forall b'\in\mcal B,\ c_{b',b}\cdot c_{b,b'}\cong id_{b\otimes b'}\}.\label{Z2center}
\end{equation}
An element of $Z_2(\mcal B)$ is called transparent. If the symmetric center is trivial, $Z_2(\mcal B)=\{1\}$, $\mcal B$ is called non-degenerate.\footnote{The non-degeneracy is equivalent to the non-degeneracy of the $S$-matrix defined by
\begin{equation}
    \widetilde S_{i,j}:=\tr(c_{b_j,b_i}\cdot c_{b_i,b_j}).\label{Smatrix}
\end{equation}
Its components
\begin{equation}
    d_i:=\widetilde S_{1,i}\label{quantumdim}
\end{equation}
are called quantum dimensions of $b_i$'s. A normalized $S$-matrix is defined by
\[ S_{i,j}:=\frac{\widetilde S_{i,j}}{\widetilde S_{1,1}}. \]
Relatedly, another modular matrix $T$ is defined by
\[ T_{i,j}:=e^{2\pi ih_i}\delta_{i,j}, \]
where $h_i$ is a conformal dimension in CFTs. They define central charge $c(\mcal B)$ (mod 8) by
\begin{equation}
    (ST)^3=e^{\pi ic(\mcal B)/4}S^2.\label{centralcharge}
\end{equation}} A non-degenerate pre-modular fusion category is called modular fusion category (MFC), or modular.

Finally, we introduce our main characters, algebra and module category. An algebra in a fusion category $\mcal C$ is a monoid. Explicitly, it is a triplet $(A,\mu,u)$ where $A\in\mcal C$ is an object, $\mu:A\otimes A\to A$ and $u:1\to A$ are morphisms subject to coherence conditions. An algebra is called connected if $\dim_\Bbbk\mcal C(1,A)=1$. A right $A$-module is a pair $(m,p)$ where $m\in\mcal C$ and $p:m\otimes A\to m$ is a morphism subject to coherence conditions. Right $A$-modules in $\mcal C$ form a category $\mcal C_A$ (some write $\text{Mod}_{\mcal C}(A)$). In our setup, $\mcal C_A$ is known to be a finite abelian category \cite{EGNO15}. A left $A$-modules and their category are defined in the same way. An algebra is called separable if the category of $A$-modules are semisimple.\footnote{For more general monoidal category $\mcal C$, separability is defined via splitting, while for fusion category, it is known \cite{O01} that the separability is equivalent to semisimplicity of $\mcal C_A$ (or $_A\mcal C$).} Let $\mcal B$ be a braided monoidal category with braiding $c$. An algebra $A\in\mcal B$ is called commutative if $\mu=\mu\cdot c_{A,A}$. An algebra is called étale if it is both commutative and separable. Note that the identity object $1\in\mcal B$ is always an (étale) algebra. It gives $\mcal B_A\simeq\mcal B$. A BFC is called completely anisotropic if there are no nontrivial étale algebras. Any étale algebra canonically decomposes as a direct sum of connected ones \cite{DMNO10}. Therefore, in classifying étale algebras in pre-modular categories, it is enough to consider connected ones. One essential fact for our purposes is the following\newline

\textbf{Theorem.} \cite{KO01,ENO02,DMNO10} \textit{Let $\mcal B$ be a BFC and $A\in\mcal B$ a connected étale algebra. Their Frobenius-Perron dimensions obey}
\begin{equation}
    \fp(\mcal B_A)=\frac{\fp(\mcal B)}{\fp_{\mcal B}(A)}.\label{FPdimBA}
\end{equation}\newline
Thanks to (\ref{FPge1}), $\fp(\mcal B_A)$ is nonzero. We can thus rewrite it as
\begin{equation}
    \fp_{\mcal B}(A)=\frac{\fp(\mcal B)}{\fp(\mcal B_A)}.\label{FPdimA}
\end{equation}
Before we explain the last piece, module category, let us introduce a subcategory of $\mcal B_A$. For a BFC $\mcal B$ and an algebra $A\in\mcal B$, an $A$-module $(m,p)$ obeying $p\cdot c_{m,A}\cdot c_{A,m}=p$ is called dyslectic (or local) \cite{P95}. The full subcategory of dyslectic modules are denoted $\mcal B_A^0\subset\mcal B_A$. It was shown in the paper that for a connected étale algebra $A\in\mcal B$, the category of dyslectic modules $\mcal B_A^0$ is a BFC. Now, we introduce module category. Let $\mcal C$ be a monoidal category. A left $\mcal C$-module category (or left module category over $\mcal C$) is a quadruple $(\mcal M,\triangleright,m,l)$ of a category $\mcal M$, a bifunctor $\triangleright:\mcal C\times\mcal M\to\mcal M$, a natural isomorphism $m_{-,-,-}:(-\otimes-)\triangleright-\cong-\triangleright(-\triangleright-)$, and another natural isomorphism $l:1\triangleright\mcal M\simeq\mcal M$ subject to coherence conditions. Right $\mcal C$-module categories are defined analogously. Let $\mcal M_1,\mcal M_2$ be (left) $\mcal C$-module categories. $\mcal M\simeq\mcal M_1\oplus\mcal M_2$ is called the direct sum of module categories $\mcal M_{1,2}$. A (left) $\mcal C$-module category $\mcal M$ is called indecomposable if it is not equivalent to a nontrivial direct sum of module categories. In discussing physical implications, we assume $\mcal M$ be finite and indecomposable. We then have the following\newline

\textbf{Theorem.} \cite{O01,EGNO15} \textit{Let $\mcal C$ be a finite multitensor category. For any finite $\mcal C$-module category $\mcal M$, $\exists A\in\mcal C$ such that $\mcal M\simeq\mcal C_A$.}\newline

\hspace{-17pt}Therefore, with the finiteness assumption, the classification of such $\mcal B$-module categories reduces to the classification of algebra objects $A\in\mcal B$. For this classification, we can make the most of (\ref{FPdimA}). Furthermore, assuming $\mcal M$ be indecomposable, we can address our question on SSB. Since all states are in ground states, in gapped phases, an SSB is equivalent to an existence of charged operator. Recalling the correspondence (\ref{1to1}), we arrive a category-theoretical\newline

\textbf{Definition.} Let $\mcal C$ be a fusion category and $\mcal M$ be a (left) $\mcal C$-module category describing a $\mcal C$-symmetric gapped phase. A symmetry $c\in\mcal C$ is called spontaneously broken if $\exists m\in\mcal M$ such that $c\triangleright m\not\cong m$. We also say $\mcal C$ is spontaneously broken if there exists a spontaneously broken object $c\in\mcal C$. A categorical symmetry $\mcal C$ is called preserved (i.e., not spontaneously broken) if all objects act trivially.\newline

\hspace{-17pt}The definition leads to the\newline

\textbf{Lemma.} \textit{Let $\mcal C$ be a fusion category and $\mcal M$ be an indecomposable (left) $\mcal C$-module category. Then, $\rank(\mcal M)>1$ implies SSB of $\mcal C$ (i.e., $\mcal C$ is spontaneously broken).}\newline

\textit{Proof.} Assume the opposite. Pick $m\in\mcal M$. By assumption, $\forall c\in\mcal C$, $c\triangleright m\cong m$. This means $m$ is a $\mcal C$-module category with rank one. This contradicts $\mcal M$ being indecomposable. $\square$\newline

\textbf{Remark.} On the other hand, if a (left) $\mcal C$-module category is \textit{not} indecomposable, one can have $\text{GSD}>1$ and preserved $\mcal C$ simultaneously. For example, let $\mcal C$ be a fusion category with rank one module category $\mcal M_1$. Pick two of them and construct $\mcal M\simeq\mcal M_1\oplus\mcal M_1$. A gapped phase described by $\mcal M$ has $\text{GSD}=2$ while $\mcal C$ is preserved because $\forall m\in\mcal M$, $\mcal C\triangleright m\cong m$. This is what we meant in the footnote \ref{directsumtheory}.\newline

Before we close this subsection, let us review one technical tool to study module categories. Let $\mcal C$ be a semisimple rigid monoidal category and $\mcal M$ be a semisimple left $\mcal C$-module category. For $m_1,m_2\in\mcal M$, the internal Hom $\Hom(m_1,m_2)$ from $m_1$ to $m_2$ is defined, if exists, by a natural isomorphism (or equivalently universality)
\begin{equation}
    \forall c\in\mcal C,\quad\mcal M(c\triangleright m_1,m_2)\cong\mcal C(c,\Hom(m_1,m_2)).\label{inthom}
\end{equation}
It has a natural isomorphism
\begin{equation}
    \forall c\in\mcal C,\forall m_1,m_2\in\mcal M,\quad\Hom(m_1,c\triangleright m_2)\cong c\otimes\Hom(m_1,m_2).\label{inthomprop}
\end{equation}
We will use this fact to translate actions $\triangleright$ of $\mcal C$ on $\mcal M$ to fusion products $\otimes$ in $\mcal C$. For every $m\in\mcal M$, $\Hom(m,m)$ has a canonical structure of an algebra in $\mcal C$. Furthermore, it is also known \cite{O01,EGNO15} that the functor
\begin{equation}
    F:=\Hom(m,-):\mcal M\to\mcal C_{\Hom(m,m)}\quad(m\in\mcal M)\label{Ffunc}
\end{equation}
sending $m'\in\mcal M$ to $\Hom(m,m')\in\mcal C$ gives $\mcal M\simeq\mcal C_{\Hom(m,m)}$.

\subsection{Method}
In this subsection, we explain our method. Recall that we take pre-modular category $\mcal B$ as our ambient category without much loss of generality, and we study finite $\mcal B$-module categories $\mcal M$'s. Then, since such module categories are equivalent to a category of $A$-modules for some algebra $A\in\mcal B$, classification of $\mcal B$-module categories reduces to classification of algebras. This is the actual problem we are going to tackle. How we classify algebras? Combining (\ref{FPge1}) and (\ref{FPdimA}), we obtain inequalities
\begin{equation}
    (1\le)\fp(\mcal B_A)\le\fp(\mcal B).\label{FPdimineq}
\end{equation}
Note that the Frobenius-Perron dimension of an unknown category $\mcal B_A$ is bounded from above by a known Frobenius-Perron dimension $\fp(\mcal B)$. The inequality leads to our main\newline

\textbf{Theorem.} \textit{Let $\mcal B$ be a BFC and $A\in\mcal B$ a connected étale algebra. An upper bound on} $\rank(\mcal B_A)$ \textit{is given by}
\begin{equation}
    \rank(\mcal B_A)\le\lfloor\fp(\mcal B)\rfloor.\label{boundrankBA}
\end{equation}\newline

\textit{Proof}. Our goal is to maximize the rank of $\mcal B_A$. To achieve that, we can tune two numbers, $\fp(\mcal B_A)$ and Frobenius-Perron dimensions of objects. First, if we increase $\fp(\mcal B_A)$, we can make ranks larger. Second, by minimizing Frobenius-Perron dimensions of objects in $\mcal B_A$, we can maximize the rank. The first is bounded from above by $\fp(\mcal B)$, and the latter is bounded from below by $\fp_{\mcal B_A}=1$. In this case, the rank is maximized, and it is given by $\lfloor\fp(\mcal B)\rfloor$.

This can be visualized as follows. Imagine a box and balls. The `size' of the box corresponds to the Frobenius-Perron dimension and the number of balls corresponds to the rank. Our goal is to maximize the number of balls. In general, the number becomes larger if the box is larger. The size of the box, i.e., $\fp(\mcal B_A)$, is bounded from above by $\fp(\mcal B)$. The number of balls is further increased by taking the balls as small as possible, which is one. When all balls have size one, i.e., pointed, the Frobenius-Perron dimension of such category coincides with the number of balls (i.e., rank). Since the number is a natural number, it is given by $\lfloor\fp(\mcal B)\rfloor$. $\square$\newline

The assumption of finiteness imply $\mcal M\simeq\mcal B_A$ be semisimple, and hence the algebra $A$ is separable. We further assume $A$ be commutative, i.e., étale. In classifying étale algebras, it is enough to study connected ones. Therefore, our problem reduces to classify connected étale algebras. Since ranks are bounded from above, our theorem leaves only finitely many candidates, and makes the classification problem tractable. Our procedure consists of three steps:
\begin{enumerate}
    \item Find a maximal rank $r_\text{max}$,
    \item List up candidate fusion categories,
    \item Check which of them satisfy axioms.
\end{enumerate}

The first step is easy; it is given by (\ref{boundrankBA})
\begin{equation}
    r_\text{max}=\lfloor\fp(\mcal B)\rfloor.\label{rmax}
\end{equation}
In the second step, we look at fusion categories\footnote{The category of right $A$-modules further becomes spherical if the ambient BFC $\mcal B$ is balanced and $A$ is a rigid connected commutative algebra with trivial twist \cite{KO01}. However, in our examples, we do not need to use this fact.} whose ranks are no larger than $\lfloor\fp(\mcal B)\rfloor$. Here, ideally, we need the complete list of fusion categories. However, such a list is known only for small ranks (up to rank two) \cite{O02}. For larger ranks, the known lists assume additional conditions such as pivotal structure (rank three) \cite{O13}, or multiplicity-free fusion rings (up to rank nine) \cite{LPR20,VS22}, summarized in the AnyonWiki \cite{anyonwiki}.\footnote{For further partial classifications, see also \cite{L14} (rank four pseudo-unitary fusion category with two self-dual simple objects), or \cite{DZD16} (non-trivially graded self-dual fusion category with rank four). We thank Sebastien Palcoux for bringing these papers to our attention.} Although our method should provide complete classification of connected étale algebras, due to the lack of complete candidates for $\mcal B_A$, we practically have to \textit{assume} $\mcal B_A$ be multiplicity-free. We emphasize that once a complete list of fusion categories are obtained, we can relax the assumption, and get the full classification. While the assumption is mathematically unsatisfactory, we believe it is physically not a serious problem for the following reason. In case $\mcal B$ is non-degenerate (i.e., modular), the categorical dimension $D(\mcal B)$ contributes free energy \cite{KP05}
\[ F\ni T\ln D(\mcal B), \]
where the temperature $T$ is given by length $1/T$ of the Euclidean time compactified to a circle. If $\mcal B$ is pseudo-unitary, we can write it $F\ni\frac12T\ln\fp(\mcal B)$. Therefore, the physical principle of minimal free energy prefers smaller Frobenius-Perron dimensions.\footnote{The same reasoning was used to explain which emergent symmetries appear in IR \cite{KK22free}.} Now, we ask this question. Let $\mcal C_1$ and $\mcal C_2$ be fusion categories with the same rank. Suppose $\mcal C_{1,2}$ have fusion rings with and without multiplicities, respectively. Which fusion categories have smaller Frobenius-Perron dimensions? The answer is $\mcal C_2$, the one without multiplicity. This is because multiplicities increase Frobenius-Perron dimensions of simple objects, and hence that of the category. Therefore, we physically expect $\mcal B_A$'s describing IR behaviors would be given by those without multiplicities. Supported by this physical argument, we assume $\mcal B_A$'s be multiplicity-free below.

The candidate fusion categories $\mcal C$'s should obey three conditions: i) $\rank(\mcal C)\le r_\text{max}$, ii) $\fp(\mcal C)\le\fp(\mcal B)$, and iii) it solves (\ref{FPdimA}). Since an algebra consists of objects in the ambient category $\mcal B$, we know their Frobenius-Perron dimensions. The formula (\ref{FPdirectsum}) gives Frobenius-Perron dimensions of algebra objects as linear sums. Then, it imposes nontrivial constraints, and rules out some fusion categories via (\ref{FPdimA}). In the final step, we study which of the remaining candidates satisfy axioms.

In checking axioms, we need $c_{A,A}$. It is computed as follows. For $A\cong a_1\oplus a_2\oplus\cdots\oplus a_n$ with product projections $p_i:A\to a_i$ and coproduct injections $\iota_i:a_i\to A$, chasing the commuting diagrams
\begin{figure}[H]
\begin{center}
\includegraphics[width=0.8\textwidth]{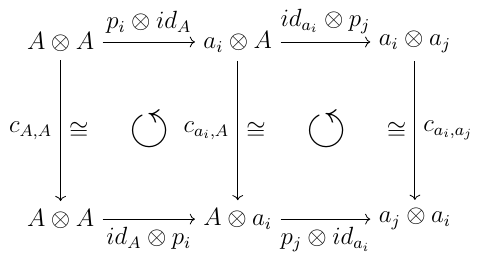},
\end{center}\caption{Computation of $c_{A,A}$}\label{computecAA}
\end{figure}
\hspace{-17pt}one finds\footnote{More generally, for $b\cong b_1\oplus\cdots\oplus b_m,b'\cong b_1'\oplus\cdots\oplus b'_n$ with product projections $p_i:b\to b_i,q_j:b'\to b'_j$ and coproduct injections $\iota_i:b_i\to b,\kappa_j:b'_j\to b'$, one has
\begin{equation}
    c_{b,b'}\cong\sum_{i=1}^m\sum_{j=1}^n(\kappa_j\otimes\iota_i)\cdot c_{b_i,b'_j}\cdot(p_i\otimes q_j).\label{cbb'}
\end{equation}}
\begin{equation}
    c_{A,A}\cong\sum_{i,j=1}^n(\iota_j\otimes\iota_i)\cdot c_{a_i,a_j}\cdot(p_i\otimes p_j).\label{cAA}
\end{equation}
Namely, it is basically a sum of braidings $c_{a_i,a_j}$ for all pairs $(a_i,a_j)$ in $A$. In order to make this abstract expression concrete, let us compute one example. Let $A\cong1\oplus\eta$ be an algebra with $\eta$ a $\mbb Z_2$ object and $c_{\eta,\eta}\cong id_1$. Then, we have
\[ A\otimes A\cong(1\oplus\eta)\otimes(1\oplus\eta)\cong1\oplus\eta\oplus\eta\oplus1. \]
The formula (\ref{cAA}) gives
\begin{align*}
    c_{A,A}\cong&(\iota_1\otimes\iota_1)\cdot c_{1,1}\cdot(p_1\otimes p_1)+(\iota_\eta\otimes\iota_1)\cdot c_{1,\eta}\cdot(p_1\otimes p_\eta)\\
    &~~+(\iota_1\otimes\iota_\eta)\cdot c_{\eta,1}\cdot(p_\eta\otimes p_1)+(\iota_\eta\otimes\iota_\eta)\cdot c_{\eta,\eta}\cdot(p_\eta\otimes p_\eta)\\
    \cong&(\iota_1\otimes\iota_1)\cdot id_1\cdot(p_1\otimes p_1)+(\iota_\eta\otimes\iota_1)\cdot id_\eta\cdot(p_1\otimes p_\eta)\\
    &~~+(\iota_1\otimes\iota_\eta)\cdot id_\eta\cdot(p_\eta\otimes p_1)+(\iota_\eta\otimes\iota_\eta)\cdot id_1\cdot(p_\eta\otimes p_\eta)\cong id_{A\otimes A}.
\end{align*}
Such an algebra is commutative because $\mu=\mu\cdot id_{A\otimes A}$.

In practice, it is tedious to compute half-braidings $c_{a_i,a_j}$'s. While it can be easily computed when the simple objects are invertible via the well-known technique \cite{K05}, one in general has to solve hexagon equations. Hence, we take an indirect path to check commutativity. If $\mu\cdot c_{A,A}=\mu$, by substituting the RHS to the LHS, one obtains a necessary condition
\begin{equation}
    \mu\cdot c_{A,A}\cdot c_{A,A}=\mu.\label{muccmu}
\end{equation}
The double braiding $c_{A,A}\cdot c_{A,A}$ is also given by the sum of simple ones:
\begin{equation}
    c_{A,A}\cdot c_{A,A}\cong\sum_{i,j=1}^n(\iota_i\otimes\iota_j)\cdot c_{a_j,a_i}\cdot c_{a_i,a_j}\cdot(p_i\otimes p_j).\label{cAAcAA}
\end{equation}
The double-braidings can be easily computed using the formula
\begin{equation}
    c_{a_j,a_i}\cdot c_{a_i,a_j}\cong\bigoplus_k {N_{ij}}^k\frac{e^{2\pi ih_k}}{e^{2\pi i(h_i+h_j)}}id_k,\label{doublebraiding}
\end{equation}
where $h_i$'s are conformal dimensions in RCFTs. With this formula, one can easily check whether the necessary condition (\ref{muccmu}) is satisfied; if it is not met, then we can conclude $A\in\mcal B$ is not commutative.

This method works for generic pre-modular category $\mcal B$ including degenerate ones. However, if $\mcal B$ is non-degenerate (i.e., modular), we can employ more constraints. Some useful facts are known as the\newline

\textbf{Theorem.} \cite{KO01,DMNO10} \textit{Let $\mcal B$ be an MFC, and $A\in\mcal B$ a connected étale algebra. Then, $\mcal B_A^0$ is an MFC with}
\begin{equation}
\begin{split}
    \fp(\mcal B_A^0)&=\frac{\fp(\mcal B)}{(\fp_{\mcal B}(A))^2},\\
    e^{2\pi ic(\mcal B)/8}&=e^{2\pi ic(\mcal B_A^0)/8}.
\end{split}\label{constsMFCB}
\end{equation}\newline

\hspace{-17pt}In particular, the matching of (additive) central charges (mod 8) turns out to be strong enough to rule out almost all candidates. We summarize our setup and some facts in the following
\begin{table}[H]
\begin{center}
\begin{tabular}{c|c|c}
    Symbol&Setup&Facts\\\hline
    $\mcal B$&Pre-modular fusion category&\\
    $A$&Connected étale algebra $A\in\mcal B$&\\
    $\mcal B_A$&Category of right $A$-modules&FC\\
    $\mcal B_A^0$&Category of dyslectic right $A$-modules&Pre-modular FC (especially MFC for modular $\mcal B$)
\end{tabular}.
\end{center}\caption{Categories and algebras}\label{catalg}
\end{table}

\subsection{Examples}
In order to demonstrate our method, we work out concrete examples\footnote{We basically follow notations in \cite{FMS}.} in this subsection. We start from non-degenerate pre-modular ambient categories (i.e., MFCs) to see our method reproduces known results. Then, we relax the non-degeneracy of braiding. The classification of connected étale algebras in degenerate pre-modular categories would be new.

\subsubsection{Modular $\mcal B$}
We first assume ambient pre-modular categories $\mcal B$'s are non-degenerate, i.e., modular. Since classification of connected étale algebras in (or module categories over) $\mcal B$ are well-known (at least for pseudo-unitary cases), we study only a few examples.\newline

\textbf{Example 1:} \bm{$M(5,4)+\phi_{1,3}.$}\newline
As a first example, we consider the tricritical Ising model. The model is described by a rank six pseudo-unitary MFC. Its $\phi_{1,3}$-deformation preserves rank three modular fusion subcategory $\mcal B$. Its simple objects are given by $\{1_0,\eta_{\frac32},N_{\frac7{16}}\}$\footnote{We write conformal dimensions in subscripts for a reader's convenience.} obeying the fusion rules of an Ising category:
\[ \forall b\in\mcal B,\quad1\otimes b\cong b\cong b\otimes1,\quad \eta\otimes\eta\cong1,\quad N\otimes N\cong 1\oplus\eta. \]
They have
\[ \fp_{\mcal B}(1)=1=\fp_{\mcal B}(\eta),\quad\fp_{\mcal B}(N)=\sqrt2, \]
and hence
\[ \fp(\mcal B)=4. \]
In the ambient category, they have modular $S$- and $T$-matrices (in the basis above)
\[ S=\frac12\begin{pmatrix}1&1&\sqrt2\\1&1&-\sqrt2\\\sqrt2&-\sqrt2&0\end{pmatrix},\quad T=\begin{pmatrix}1&&\\&-1&\\&&e^{7\pi i/8}\end{pmatrix}, \]
and hence central charge (\ref{centralcharge})
\begin{equation}
    c(\mcal B)=\frac72\quad(\text{mod }8).\label{cBtri}
\end{equation}
In this case, our upper bound is given by
\[ r_\text{max}=\lfloor\fp(\mcal B)\rfloor=4. \]
We thus search for an MFC with rank no larger than four and central charge (\ref{cBtri}) mod 8. Fortunately, MFCs with small ranks are completely classified. According to \cite{GK94,rank4}, there is a unique unitary MFC satisfying the conditions. It has $SU(2)_2$ realization and central charge $\frac72$ mod 4. Since the MFC has Frobenius-Perron dimension $\fp(\mcal B_A^0)=4=\fp(\mcal B)$, the full category of (right) $A$-modules cannot have more simple objects. We conclude
\begin{equation}
    \mcal B_A^0\simeq\mcal B_A\simeq\mcal B.\label{BAtri}
\end{equation}
This result together with the dimension formula (\ref{FPdimA}) implies
\begin{equation}
    \fp_{\mcal B}(A)=1.\label{dimAtri}
\end{equation}
In other words, $\mcal B$ is completely anisotropic as experts know. Physically, the result implies
\begin{equation}
    \text{GSD}=3\label{GSDtri}
\end{equation}
in $\mcal B$-symmetric gapped phases if we assume finiteness of module categories. We will see numerical results in section \ref{numericalcheck} is consistent with the mathematical result. This implies $\mcal B$ is spontaneously broken. Since classification of connected étale algebras in pseudo-unitary MFCs are well-known, we next relax pseudo-unitarity.\newline

\textbf{Example 2:} \bm{$M(3,5)+\phi_{1,2}.$}\newline
Next, let us study examples with non-pseudo-unitary modular ambient categories. We would like to demonstrate our method works even in these cases. As a first example, pick the non-unitary minimal model $M(3,5)$ as our UV theory. Its $\phi_{1,2}$-deformation preserves a rank two non-pseudo-unitary MFC $\mcal B$. Its simple objects are given by $\{1_0,W_{\frac15}\}$ obeying the fusion rules of a Fibonacci category:
\[ \forall b\in\mcal B,\quad1\otimes b\cong b\cong b\otimes1,\quad W\otimes W\cong1\oplus W. \]
Thus, they have
\[ \fp_{\mcal B}(1)=1,\quad\fp_{\mcal B}(W)=\zeta, \]
where $\zeta:=\frac{1+\sqrt5}2$. In the ambient category, they have modular $S$- and $T$-matrices (in the basis above)
\[ S=\pm\frac{\sqrt2}{\sqrt{5-\sqrt5}}\begin{pmatrix}1&-\zeta^{-1}\\-\zeta^{-1}&-1\end{pmatrix},\quad T=\begin{pmatrix}1&\\&e^{2\pi i/5}\end{pmatrix}, \]
and hence $\mcal B$ has central charge (\ref{centralcharge})
\begin{equation}
    c(\mcal B)=\begin{cases}\frac25&(D>0)\\-\frac{18}5&(D<0)\end{cases}\quad(\text{mod }8).\label{cBM35}
\end{equation}
Since we are not assuming $\mcal B$ be pseudo-unitary, the categorical dimension $D(\mcal B)$ can be negative. This leads to the two possibilities. Since the ambient category has Frobenius-Perron dimension
\[ \fp(\mcal B)=\frac{5+\sqrt5}2\approx3.6, \]
the maximum rank of $\mcal B_A$ is three
\begin{equation}
    r_\text{max}=\lfloor\fp(\mcal B)\rfloor=3.\label{rmaxM35}
\end{equation}
Luckily, MFCs up to rank three are completely classified. According to \cite{GK94}, we can rule out all but one fusion ring with realization $PSU(2)_3$. Since both $\mcal B$ and the putative MFC have Frobenius-Perron dimension $\frac{5+\sqrt5}2$, the only allowed algebra is $A\cong1$. The trivial algebra gives the regular module category $\mcal M\simeq\mcal B\simeq\mcal B_A$. Physically, with the finiteness assumption, the result implies
\begin{equation}
    \text{GSD}=2,\label{GSDM35+12}
\end{equation}
and $\mcal B$ is spontaneously broken.\newline

\textbf{Example 3:} $M(4,7)+\phi_{1,2}.$\newline
As a final modular example, we take the non-unitary minimal model $M(4,7)$ as our UV theory. Its $\phi_{1,2}$-deformation preserves rank three non-degenerate pre-modular category 
\[ \mcal B=\{1_0,(\mcal L_{5,1})_{\frac{10}7},(\mcal L_{3,1})_{\frac17}\}. \]
They have monoidal products
\begin{table}[H]
\begin{center}
\begin{tabular}{c|c|c|c}
    $\otimes$&1&$\mcal L_{5,1}$&$\mcal L_{3,1}$\\\hline
    1&1&$\mcal L_{5,1}$&$\mcal L_{3,1}$\\\hline
    $\mcal L_{5,1}$&&$1\oplus\mcal L_{3,1}$&$\mcal L_{5,1}\oplus\mcal L_{3,1}$\\\hline
    $\mcal L_{5,1}$&&&$1\oplus\mcal L_{5,1}\oplus\mcal L_{3,1}$
\end{tabular},
\end{center}
\end{table}
\hspace{-17pt}and hence Frobenius-Perron dimensions
\[ \fp_{\mcal B}(1)=1,\quad\fp_{\mcal B}(\mcal L_{5,1})=\frac{\cos\frac{3\pi}{14}}{\sin\frac\pi7}\approx1.802,\quad\fp_{\mcal B}(\mcal L_{3,1})=1+2\sin\frac{3\pi}{14}\approx2.247. \]
Note that since the ambient category is non-pseudo-unitary, they are different from quantum dimensions\footnote{The categorical dimension is $D(\mcal B)^2=\frac7{4\cos^2\frac\pi{14}}$.}
\[ d_1=1,\quad d_{5,1}=2\sin\frac\pi{14},\quad d_{3,1}=-\frac1{2\sin\frac{3\pi}{14}}. \]
The Frobenius-Perron dimension of the ambient category is given by
\[ \fp(\mcal B)=\frac7{4\sin^2\frac\pi7}\approx9.3. \]
Thus, an upper bound on ranks of $\mcal B_A$'s is
\[ r_\text{max}=9. \]
In order to classify $\mcal B_A$, let us first find $\mcal B_A^0$. To achieve the goal, we first compute central charge of the ambient category. The modular $S$- and $T$-matrices are given as (in the basis above)
\[ S=\pm\frac{2\cos\frac\pi{14}}{\sqrt7}\begin{pmatrix}1&2\sin\frac\pi{14}&-\frac1{2\sin\frac{3\pi}{14}}\\2\sin\frac\pi{14}&\frac1{2\sin\frac{3\pi}{14}}&1\\-\frac1{2\sin\frac{3\pi}{14}}&1&-2\sin\frac\pi{14}\end{pmatrix},\quad T=\begin{pmatrix}1&&\\&e^{6\pi i/7}&\\&&e^{2\pi i/7}\end{pmatrix}. \]
Therefore, it has central charge\footnote{We get
\[ (ST)^3=\pm e^{\pi i/7}S^2. \]
Solving $\pm e^{\pi i/7}=e^{\pi i c/4}$ gives the (additive) central charge.}
\[ c(\mcal B)=\frac47,-\frac{24}7\quad(\text{mod }8), \]
respectively. Which MFC can be the category $\mcal B_A^0$ of dyslectic modules? In the list \cite{GK94}, the only fusion ring which can match the central charge has rank three with $PSU(2)_5$ realization. Since the MFC has $\fp(\mcal C)=\frac7{4\sin^2\frac\pi7}$, the corresponding algebra should have $\fp_{\mcal B}(A)=1$. Taking into account the Frobenius-Perron dimensions of simple objects, the only possibility is
\[ A\cong1. \]
The trivial algebra gives the regular module category
\begin{equation}
    \mcal B_A^0\simeq\mcal B_A\simeq\mcal B.\label{BA47}
\end{equation}
We have learned this is the only module category up to rank six. However, our upper bound allows module categories up to rank nine. Hence, we also have to study candidates with ranks $7,8,9$. \cite{rank9} gave a partial list of (unitary) modular categories up to rank nine. We find none of them with ranks $7,8,9$ can match the central charge of the ambient category. This suggests the only connected étale algebra is the trivial one (or $\mcal B$ is completely anisotropic). In order to prove this statement, we need a full list of modular categories up to rank nine. Physically, the result suggests
\begin{equation}
    \text{GSD}=3,\label{GSDM47+12}
\end{equation}
and $\mcal B$ is spontaneously broken.

\subsubsection{Non-modular $\mcal B$}
Next, we relax the assumption of non-degeneracy. As in modular cases, we first assume pseudo-unitarity, and later further relax the assumption.\newline

\textbf{Example 4:} $M(6,5)+\phi_{1,3}.$\newline
In our first example, we pick the tetracritical Ising model $M(6,5)$ as our UV theory. Its $\phi_{1,3}$-deformation preserves pre-modular category $\mcal B$ with rank four
\[ \mcal B\simeq\fib\boxtimes\vect_{\mbb Z_2}^1. \]
The $\mbb Z_2$ object $\eta\in\vect_{\mbb Z_2}^1$ is transparent, i.e., $\eta\in Z_2(\mcal B)$. Namely, $\mcal B$ is degenerate. The four simple objects\footnote{We abbreviate $\eta\otimes W$ to $\eta W$.} $\{1_0,\eta_3,W_{\frac75},(\eta W)_{\frac25}\}$ have monoidal products
\begin{table}[H]
\begin{center}
\begin{tabular}{c|c|c|c|c}
    $\otimes$&1&$\eta$&$W$&$\eta W$\\\hline
    1&1&$\eta$&$W$&$\eta W$\\\hline
    $\eta$&&1&$\eta W$&$W$\\\hline
    $W$&&&$1\oplus W$&$\eta\oplus\eta W$\\\hline
    $\eta W$&&&&$1\oplus W$
\end{tabular},
\end{center}
\end{table}
\hspace{-17pt}and Frobenius-Perron dimensions
\[ \fp_{\mcal B}(1)=1=\fp_{\mcal B}(\eta),\quad\fp_{\mcal B}(W)=\zeta=\fp_{\mcal B}(\eta W). \]
Thus, $\mcal B$ has
\[ \fp(\mcal B)=5+\sqrt5\approx7.2. \]
The Frobenius-Perron dimension gives an upper bound on ranks of $\mcal B_A$'s
\[ r_\text{max}=7. \]
Now, our classification problem reduced to find fusion categories $\mcal C$'s with ranks no larger than seven, and which solve
\[ \fp_{\mcal B}(A)=m+\zeta n=\frac{5+\sqrt5}{\fp(\mcal C)} \]
with $m,n\in\mbb N$.

Given an upper bound on ranks of $\mcal B_A$'s, let us rule out some of them. First, we can rule out those with odd ranks. This was recently used to show an RG flow with emergent supersymmetry \cite{KKSUSY}. (Also see references therein.) A skeptical reader can try to find NIM-reps. For example, the absence of rank one $\mcal B_A$ can be easily shown by realizing that there is no natural number $n_W$ solving
\[ n_W^2=1+n_W. \]
We are thus left with ranks two, four, and six.

In order to list up candidates for $\mcal B_A$, we scan lists of fusion categories up to rank six \cite{O02,O13,LPR20,anyonwiki}. We find only three fusion rings
\[ \text{FR}^{2,0}_1,\text{FR}^{2,0}_2,\text{FR}^{4,0}_2 \]
satisfy the conditions. (Note that we can rule out all rank six fusion rings employing (\ref{FPdimA}) or $\fp(\mcal C)\le\fp(\mcal B)=5+\sqrt5\approx7.2$.) Clearly, the rank four candidate is allowed; it corresponds to the trivial algebra $A\cong1$ giving $\mcal B_A\simeq\mcal B$.

The other two candidates both have rank two. To find all rank two candidates, we search for two-dimensional NIM-reps. At rank two, we find a unique (up to basis transformation) solution
\begin{equation}
    n_1=1_2=n_\eta,\quad n_W=\begin{pmatrix}1&1\\1&0\end{pmatrix}=n_{\eta W}.\label{M65+132dimNIMrep}
\end{equation}
Let $m_1,m_2\in\mcal C$ be a basis. Namely, they are representatives of simple objects. The NIM-rep means our degenerate rank four pre-modular category $\mcal B$ acts as
\begin{equation}
\begin{split}
    1\triangleright m_j&\cong m_j\cong\eta\triangleright m_j,\\
    W\triangleright m_1&\cong m_1\oplus m_2\cong\eta W\triangleright m_1,\\
    W\triangleright m_2&\cong m_1\cong\eta W\triangleright m_2.
\end{split}\label{M65+13rank2actionM}
\end{equation}
Recalling the defining natural isomorsphisms of internal Homs, the actions are translated to fusion products
\begin{equation}
\begin{split}
    F(m_j)\cong F(1\triangleright m_j)\equiv\Hom(m,1\triangleright m_j)&\cong\Hom(m,m_j),\\
    F(m_j)\cong F(\eta\triangleright m_j)\equiv\Hom(m,\eta\triangleright m_j)&\cong\eta\otimes\Hom(m,m_j),\\
    F(m_1)\oplus F(m_2)\cong F(W\triangleright m_1)\equiv\Hom(m,W\triangleright m_1)&\cong W\otimes\Hom(m,m_1),\\
    F(m_1)\cong F(W\triangleright m_2)\equiv\Hom(m,W\triangleright m_2)&\cong W\otimes\Hom(m,m_2),\\
    \text{the last two natural isomorphisms with }W&\text{ replaced by }\eta W.
\end{split}\label{M65+13rank2actionA}
\end{equation}
Recall $A\cong F(m)$. In order to find its form, we set an ansatz
\[ A\cong F(m)\cong a1\oplus b\eta\oplus cW\oplus d\eta W \]
with $a,b,c,d\in\mbb N$, and perform case analysis. If $m\cong m_1$, the second fusion product demands
\[ a=b,\quad c=d. \]
For the algebra to be connected, we also need $a=1$. Thus, the ansatz reduces to $A\cong 1\oplus\eta\oplus cW\oplus c\eta W$. The third fusion product gives
\[ F(m_2)\cong(c-1)1\oplus(c-1)\eta\oplus W\oplus\eta W. \]
The other fusion products do not impose further constraints. Now, the candidate has
\[ \fp_{\mcal B}(A)=2+2c\zeta. \]
This can only solve (\ref{FPdimA}) for $\text{FR}^{2,0}_2$ with $c=0$:
\begin{equation}
    A\cong F(m_1)\cong1\oplus\eta.\label{M65+13rank2alg}
\end{equation}
The same analysis for the other case $m\cong m_2$ also leads to $A\cong1\oplus\eta$. Thus, we find the only rank two $\mcal B_A$ is $\fib$ whose fusion ring is given by $\text{FR}^{2,0}_2$. The algebra corresponds to gauge the $\mbb Z_2$ subcategory of $\mcal B$. Since we know the $\mbb Z_2$ is anomaly-free, the algebra does exist. Furthermore, the algebra is commutative because $c_{A,A}\cong id_{A\otimes A}$ thanks to $c_{\eta,\eta}\cong id_1$. It is also connected.

To summarize, we found two connected étale algebras\footnote{Since $\mcal B_A$'s are fusion categories, separability of $A$'s are automatic.} (assuming multiplicity-free fusion rings)
\begin{table}[H]
\begin{center}
\begin{tabular}{c|c|c}
    Connected étale algebra $A$&$\mcal B_A$&$\rank(\mcal B_A)$\\\hline
    $1$&$\mcal B$&4\\
    $1\oplus\eta$&$\fib$&2
\end{tabular}.
\end{center}\caption{Results on $M(6,5)+\phi_{1,3}$}\label{M65+13results}
\end{table}
Let us comment on physical implications. If we assume finiteness of $\mcal B$-module categories $\mcal M$'s, we have $\mcal M\simeq\mcal B_A$. The result tells us
\begin{equation}
    \text{GSD}\in\{2,4\},\label{GSDM65+13}
\end{equation}
and signals SSB of $\mcal B$.\newline

\textbf{Example 5:} $\hat{\frak{su}}(3)_3+\phi_{\widehat{[1;1,1]}}.$\newline
Next, let us also take an example not from minimal models but from WZW models. We pick the $\hat{\frak{su}}(3)_3$ WZW model as our UV theory. Its $\phi_{\widehat{[1;1,1]}}$-deformation preserves rank three pre-modular category
\[ \mcal B=\{1_0,\eta_1,(\eta^2)_1\}. \]
They generate the $\mbb Z_3$ symmetry of the theory. They are all transparent as evident from the monodromy charge matrix\footnote{The matrix is defined as
\begin{equation}
    M_{i,j}:=\frac{S_{i,j}S_{1,1}}{S_{1,i}S_{1,j}}.\label{monodromychargematrix}
\end{equation}
It is known \cite{K05} that
\[ b\in Z_2(\mcal B)\iff\forall b'\in\mcal B,\quad M_{b,b'}=1. \]} \cite{SY90,FRS04,K05,fMTC}
\[ M=\begin{pmatrix}1&1&1\\1&1&1\\1&1&1\end{pmatrix}. \]
Since they have apparent $\mbb Z_3$ monoidal products, they all have $\fp_{\mcal B}=1$. Thus, the surviving ambient category has
\[ \fp(\mcal B)=3, \]
and $\mcal B_A$'s have ranks bounded from above by
\[ r_\text{max}=3. \]
According to \cite{anyonwiki}, there are seven multiplication-free fusion rings up to rank three. Which of them can be $\mcal B_A$'s? In order to rule out some of them, we try to solve (\ref{FPdimA}). Since all simple objects of $\mcal B$ have Frobenius-Perron dimension one, the only allowed Frobenius-Perron dimensions are one and three. The latter possibility is realized by the trivial algebra $A\cong1$ giving $\mcal B_A\simeq\mcal B$. The former possibility requires an algebra with Frobenius-Perron dimension three. Since any étale algebras are self-dual \cite{DMNO10}, there are just two possibilities, $1\oplus1\oplus1$, or $1\oplus\eta\oplus\eta^2$. The first candidate is not connected, and we are left with the second one. The second possibility gives an algebra because we know the $\mbb Z_3$ symmetry is anomaly-free, and it can be gauged. We found two connected étale\footnote{They are both commutative because they both have
\[ c_{A,A}\cong id_{A\otimes A}. \]
The other two axioms on separability and connectedness are automatic.} algebra objects (assuming multiplicity-free $\mcal B_A$)
\begin{table}[H]
\begin{center}
\begin{tabular}{c|c|c}
    Connected étale algebra $A$&$\mcal B_A$&$\rank(\mcal B_A)$\\\hline
    $1$&$\mcal B$&3\\
    $1\oplus\eta\oplus\eta^2$&$\text{Vect}_{\mbb C}$&1
\end{tabular}.
\end{center}\caption{Results on $\hat{\mfrak{su}}(3)_3+\phi_{\widehat{[1;1,1]}}$}\label{su(3)3+adjresults}
\end{table}
The result means $\mcal B$ fails to be completely anisotropic. Physically, this result suggests the only allowed GSDs in $\mcal B$-symmetric gapped phases are one and three:
\begin{equation}
    \text{GSD}\in\{1,3\}.\label{GSDsu33}
\end{equation}
The result alone cannot say whether $\mcal B$ symmetry is spontaneously broken or not.\newline

\textbf{Example 6:} $M(6,5)+\phi_{2,1}.$\newline
In our next example, let us pick the tetracritical Ising model $M(6,5)$ as our UV theory. Its $\phi_{2,1}$-deformation preserves pre-modular category with rank three
\[ \mcal B=\{1_0,\eta_3,M_{\frac23}\}. \]
They have monoidal products
\begin{table}[H]
\begin{center}
\begin{tabular}{c|c|c|c}
    $\otimes$&1&$\eta$&$M$\\\hline
    $1$&1&$\eta$&$M$\\\hline
    $\eta$&&1&$M$\\\hline
    $M$&&&$1\oplus\eta\oplus M$
\end{tabular},
\end{center}
\end{table}
\hspace{-17pt}and hence Frobenius-Perron dimensions
\[ \fp_{\mcal B}(1)=1=\fp_{\mcal B}(\eta),\quad\fp_{\mcal B}(M)=2. \]
The $\mbb Z_2$ object $\eta$ is transparent, and $\mcal B$ is degenerate. The ambient category has Frobenius-Perron dimension
\[ \fp(\mcal B)=6. \]
Thus, an upper bound on ranks of $\mcal B_A$'s is
\[ r_\text{max}=6. \]
According to \cite{anyonwiki}, we find $\mcal B_A$ can have seven multiplication-free fusion rings
\[ \text{FR}^{1,0}_1,\text{FR}^{2,0}_1,\text{FR}^{3,2}_1,\text{FR}^{3,0}_2,\text{FR}^{4,2}_2,\text{FR}^{6,2}_1,\text{FR}^{6,4}_1. \]

Let us look at the candidates in detail. For the candidates with $\fp=6$ to be connected, the only possibility is the trivial algebra $A\cong1$ giving $\mcal B_A\simeq\mcal B$. (This observation rules out rank four and six.) For the remaining rank three candidate to be connected, the only possibility is $1\oplus\eta$. Since $\eta$ generates anomaly-free $\mbb Z_2$ symmetry, we know this algebra\footnote{One can check it is connected étale.} does exist. Hence, there is a rank three category $\mcal B_A$ of $A$-modules whose fusion ring is given by $\text{FR}^{3,2}_1$.\footnote{This is similar to anyon condensation \cite{BSS02,BSS02'} of $\eta$ although our ambient category is degenerate.} Next, in order to check an existence of rank one module category, we search for one-dimensional NIM-reps. One finds a unique solution
\begin{equation}
    n_1=1=n_\eta,\quad n_M=2.\label{M651dimNIMrep}
\end{equation}
Denoting the unique simple object $m$, we get the actions
\begin{equation}
\begin{split}
    1\triangleright m\cong&m\cong\eta\triangleright m,\\
    M\triangleright m\cong&2m.
\end{split}\label{M65+21rank1actions}
\end{equation}
The actions are translated to fusion products via the internal Homs:
\begin{equation}
\begin{split}
    F(m)\cong F(1\triangleright m)\equiv\Hom(m,1\triangleright m)\cong&\Hom(m,m),\\
    F(m)\cong F(\eta\triangleright m)\equiv\Hom(m,\eta\triangleright m)\cong&\eta\otimes\Hom(m,m),\\
    F(m)\oplus F(m)\cong F(M\triangleright m)\equiv\Hom(m,M\triangleright m)\cong&M\otimes\Hom(m,m).
\end{split}\label{M65+21rank1fusion}
\end{equation}
Recalling $A\cong\Hom(m,m)$ (if there exists rank one $\mcal B_A$) and our assumption that $A$ be connected, we get the unique candidate from the fusion products:\footnote{To find this form, start from an ansatz
\[ F(m)\cong a1\oplus b\eta\oplus cM \] with $a,b,c\in\mbb N$. Substituting the ansatz in the fusion products, one finds $a=b\& c=2a$. For $F(m)$ to be connected, we need $a=1$, and this uniquely fixes the form (\ref{M65+21rank1alg}).}
\begin{equation}
    A\cong1\oplus\eta\oplus2M.\label{M65+21rank1alg}
\end{equation}
Indeed, it has the correct Frobenius-Perron dimension $\fp_{\mcal B}(A)=6$ to solve (\ref{FPdimA}). While it is connected and separable, it fails to be commutative.\footnote{The double braiding is given by
\[ c_{A,A}\cdot c_{A,A}\cong(2+4e^{-2\pi i/3})\iota\cdot id_1\cdot p+(2+4e^{-2\pi i/3})\iota\cdot id_\eta\cdot p+(8+4e^{2\pi i/3})\iota\cdot id_M\cdot p\not\cong id_{A\otimes A}. \]} Thus, we discard the candidate. Finally, in order to see whether there exists rank two $\mcal B_A$, we search for two-dimensional NIM-reps. One finds three inequivalent (up to basis transformations) solutions
\begin{equation}
\begin{split}
    n_1=1_2,\quad n_\eta=\begin{pmatrix}0&1\\1&0\end{pmatrix},\quad&n_M=\begin{pmatrix}1&1\\1&1\end{pmatrix},\\
    n_1=1_2=n_\eta,\quad&n_M=\begin{pmatrix}0&1\\2&1\end{pmatrix},\\
    n_1=1_2=n_\eta,\quad&n_M=\begin{pmatrix}2&0\\0&2\end{pmatrix}.
\end{split}\label{M65+21rank2NIMreps}
\end{equation}
Let $m_1,m_2\in\mcal B_A$ be a basis. The first solution gives the actions
\begin{equation}
\begin{split}
    1\triangleright m_j\cong&m_j,\\
    \eta\triangleright m_1\cong&m_2,\\
    \eta\triangleright m_2\cong&m_1,\\
    M\triangleright m_j\cong&m_1\oplus m_2,
\end{split}\label{M65+21rank2action1}
\end{equation}
or fusion products
\begin{equation}
\begin{split}
    F(m_j)\cong F(1\triangleright m_j)\equiv\Hom(m,1\triangleright m_j)\cong&\Hom(m,m_j),\\
    F(m_2)\cong F(\eta\triangleright m_1)\equiv\Hom(m,\eta\triangleright m_1)\cong&\eta\otimes\Hom(m,m_1),\\
    F(m_1)\cong F(\eta\triangleright m_2)\equiv\Hom(m,\eta\triangleright m_2)\cong&\eta\otimes\Hom(m,m_2),\\
    F(m_1)\oplus F(m_2)\cong F(M\triangleright m_j)\equiv\Hom(m,M\triangleright m_j)\cong&M\otimes\Hom(m,m_j).
\end{split}\label{M65+21rank2fusion1}
\end{equation}
With an ansatz
\[ F(m_1)\cong a1\oplus b\eta\oplus cM, \]
the fusion products give
\[ c=a+b. \]
In other words, fusion products alone are not strong enough to fix the object $F(m_1)$. However, for $m_j\cong m$, we have to get an algebra $F(m)\cong\Hom(m,m)\cong A$. Without loss of generality, we write such $m$ as $m_1$. Then, in order to get connected algebra, we need $a=1$, $F(m_1)\cong A\cong1\oplus b\eta\oplus(1+b)M$. For the object to solve (\ref{FPdimA}), we need
\[ \fp_{\mcal B}(A)=3+3b\stackrel!=3, \]
or $b=0$. We arrive
\begin{equation}
    A\cong F(m_1)\cong1\oplus M,\quad F(m_2)\cong\eta\oplus M.\label{M65+21rank2alg1}
\end{equation}
The same analysis for the second and third solutions tells us they cannot give connected algebras giving rank two $\mcal B_A$'s. Thus, the only candidate for connected algebras is (\ref{M65+21rank2alg1}). While it is connected and separable, it again fails to be commutative.\footnote{The candidate $1\oplus M$ has double braiding
\[ c_{A,A}\cdot c_{A,A}\cong(1+e^{-2\pi i/3})\iota\cdot id_1\cdot p+e^{-2\pi i/3}\iota\cdot id_\eta\cdot p+(2+e^{2\pi i/3})\iota\cdot id_M\cdot p\not\cong id_{A\otimes A}. \]}

To summarize, we found
\begin{table}[H]
\begin{center}
\begin{tabular}{c|c|c}
    Connected étale algebra $A$&$\mcal B_A$&$\rank(\mcal B_A)$\\\hline
    1&$\mcal B$&3\\
    $1\oplus\eta$&$\vect_{\mbb Z_3}^1$&3
\end{tabular}.
\end{center}\caption{Results on $M(6,5)+\phi_{2,1}$}\label{M65+21results}
\end{table}
\hspace{-17pt}Since the second connected étale algebra corresponds to gauge anomaly-free $\mbb Z_2$ symmetry, the quantum $\mbb Z_3$ is anomaly-free $\alpha=1$.

Let us comment on physical implications of this result. If we assume $\mcal B$-module categories describing $\mcal B$-symmetric gapped phases are indecomposable, finite  (thus $\mcal M\simeq\mcal B_A$), has multiplicity-free fusion ring $\mcal B_A$, and $A\in\mcal B$ be connected étale, then GSD should be three:
\begin{equation}
    \text{GSD}=3.\label{GSDM65+21}
\end{equation}
This also implies $\mcal B$ symmetry should be spontaneously broken. Note that, since all simple objects have integer quantum dimensions, the Theorem 2 in \cite{TW19} does not apply here, and we cannot rule out the possibility of $\text{GSD}=1$ just from their argument. We saw a more detailed analysis could strengthen constraints on IR behaviors.

We also perform numerical checks in section \ref{numericalcheck}. Indeed, for positive Lagrangian coupling $\lambda_{2,1}>0$, we find three lowest energy eigenvalues coincide in the IR limit. The gapped phase should be described by $\mcal B$ or $\vect_{\mbb Z_3}^1$. On the other hand, for negative Lagrangian coupling $\lambda_{2,1}<0$, we find a gapped phase with $\text{GSD}=2$. Our analysis above signals the phase is described by $\mcal B_A$ with $A\cong1\oplus M$. This numerical result suggests that, in general, gapped phases with possibly degenerate pre-modular fusion category symmetries are described by non-commutative algebras.\newline

\textbf{Example 7:} $M(3,8)+\phi_{1,2}.$\newline
Finally, let us also relax the assumption of pseudo-unitarity. We pick the non-unitary minimal model $M(3,8)$ as our UV theory. Its $\phi_{1,2}$-deformation preserves rank four pre-modular category $\mcal B$:\footnote{Since $\eta$ has conformal dimension $\frac32$, the $\mbb Z_2$ algebra $1\oplus\eta$ fails to be commutative. Explicitly,
\[ c_{A,A}\cong \iota\cdot id_1\cdot p+\iota\cdot id_\eta\cdot p+\iota\cdot id_\eta\cdot p-\iota\cdot id_1\cdot p\not\cong id_{A\otimes A}. \]}
\[ \mcal B=\{1_0,\eta_{\frac32},(\mcal L_{3,1})_{-\frac14},(\mcal L_{5,1})_{\frac14}\cong\eta\otimes\mcal L_{3,1}\}. \]
They have monoidal products
\begin{table}[H]
\begin{center}
\begin{tabular}{c|c|c|c|c}
    $\otimes$&1&$\eta$&$\mcal L_{3,1}$&$\mcal L_{5,1}$\\\hline
    1&1&$\eta$&$\mcal L_{3,1}$&$\mcal L_{5,1}$\\\hline
    $\eta$&&1&$\mcal L_{5,1}$&$\mcal L_{3,1}$\\\hline
    $\mcal L_{3,1}$&&&$1\oplus\mcal L_{3,1}\oplus\mcal L_{5,1}$&$\eta\oplus\mcal L_{3,1}\oplus\mcal L_{5,1}$\\\hline
    $\mcal L_{5,1}$&&&&$1\oplus\mcal L_{3,1}\oplus\mcal L_{5,1}$
\end{tabular},
\end{center}
\end{table}
\hspace{-17pt}and
\[ \fp_{\mcal B}(1)=1=\fp_{\mcal B}(\eta),\quad\fp_{\mcal B}(\mcal L_{3,1})=1+\sqrt2=\fp_{\mcal B}(\mcal L_{5,1}). \]
The $\mbb Z_2$ object $\eta\in\mcal B$ is transparent, and $\mcal B$ is non-modular. It has
\[ \fp(\mcal B)=4(2+\sqrt2)\approx13.7. \]
Thus, ranks of $\mcal B_A$'s are bounded from above by
\[ r_\text{max}=13. \]
In the list of \cite{anyonwiki}, we find only seven fusion rings\footnote{We can discard fusion rings such as $\text{FR}^{4,2}_3,\text{FR}^{5,0}_2,\text{FR}^{7,2}_1$ which cannot be categorified to fusion categories at the outset of our search.}
\[ \text{FR}^{1,0}_1,\text{FR}^{2,0}_1,\text{FR}^{3,0}_1,\text{FR}^{4,0}_1,\text{FR}^{4,2}_1,\text{FR}^{4,0}_4,\text{FR}^{4,2}_4. \]
can solve (\ref{FPdimA}).

Let us look at the candidates in detail. We start from the last two with $\fp=\fp(\mcal B)$. For an algebra to be connected, the only possibility is $A\cong1$, and this does exist; the trivial connected étale algebra. It gives the (rank four) regular module category $\mcal M\simeq\mcal B_A\simeq\mcal B$. Next, we find the rank one candidate does not exist because there is no one-dimensional NIM-rep. (This rules out $\text{FR}^{1,0}_1$.) Thirdly, in order to find all rank two $\mcal B_A$'s, we search for two-dimensional NIM-reps. One finds a unique (up to basis transformations) solution
\begin{equation}
    n_1=1_2=n_\eta,\quad n_{31}=\begin{pmatrix}0&1\\1&2\end{pmatrix}=n_{51}.\label{M38rank2NIMrep}
\end{equation}
To figure out which rank two fusion category can be the $\mcal B$-module category, we translate the actions to fusion products. Let $m_1,m_2\in\mcal B_A$ be a basis. The solution gives the actions
\begin{equation}
\begin{split}
    1\triangleright m_j&\cong m_j\cong\eta\triangleright m_j,\\
    \mcal L_{3,1}\triangleright m_1&\cong m_2\cong\mcal L_{5,1}\triangleright m_1,\\
    \mcal L_{3,1}\triangleright m_2&\cong m_1\oplus2m_2\cong\mcal L_{5,1}\triangleright m_2.
\end{split}\label{M38rank2actions}
\end{equation}
or fusion products
\begin{equation}
\begin{split}
    F(m_j)\cong F(1\triangleright m_j)\equiv\Hom(m,1\triangleright m_j)&\cong\Hom(m,m_j)\cong\eta\otimes\Hom(m,m_j),\\
    F(m_2)\cong F(\mcal L_{3,1}\triangleright m_1)\equiv\Hom(m,\mcal L_{3,1}\triangleright m_1)&\cong\mcal L_{3,1}\otimes\Hom(m,m_1)\cong\mcal L_{5,1}\otimes\Hom(m,m_1),\\
    F(m_1)\oplus2F(m_2)\cong F(\mcal L_{3,1}\triangleright m_2)\equiv\Hom(m,\mcal L_{3,1}\triangleright m_2)&\cong\mcal L_{3,1}\otimes\Hom(m,m_2)\cong\mcal L_{5,1}\otimes\Hom(m,m_2).
\end{split}\label{M38rank2fusions}
\end{equation}
In order to solve (\ref{FPdimA}), our routine method gives
\begin{equation}
    A\cong1\oplus\eta\oplus\mcal L_{3,1}\oplus\mcal L_{5,1},\label{M38+12rank2alg}
\end{equation}
together with $m\cong m_1$ to realize $\text{FR}^{2,0}_1$. This fails to be commutative.\footnote{It has double braiding
\[ c_{A,A}\cdot c_{A,A}\cong4\iota\cdot id_{3,1}\cdot p+4\iota\cdot id_{5,1}\cdot p. \]}

Fourthly, in order to find all rank three $\mcal B_A$'s, we search for three-dimensional NIM-reps. We find two inequivalent (up to basis transformations) solutions
\begin{equation}
\begin{split}
    n_1=1_3,\quad n_\eta=\begin{pmatrix}0&0&1\\0&1&0\\1&0&0\end{pmatrix},\quad n_{31}=\begin{pmatrix}0&1&1\\1&1&1\\1&1&0\end{pmatrix},\quad n_{51}=\begin{pmatrix}1&1&0\\1&1&1\\0&1&1\end{pmatrix},\\
    n_1=1_3,\quad n_\eta=\begin{pmatrix}0&0&1\\0&1&0\\1&0&0\end{pmatrix},\quad n_{31}=\begin{pmatrix}1&1&0\\1&1&1\\0&1&1\end{pmatrix},\quad n_{51}=\begin{pmatrix}0&1&1\\1&1&1\\1&1&0\end{pmatrix}.
\end{split}\label{M38rank3NIM}
\end{equation}
Let $m_1,m_2,m_3\in\mcal B_A$ be a basis. The first solution yields  the fusion products
\begin{table}[H]
\begin{center}
\begin{tabular}{c|c|c|c}
$b\otimes\backslash$&$F(m_1)$&$F(m_2)$&$F(m_3)$\\\hline
$1$&$F(m_1)$&$F(m_2)$&$F(m_3)$\\
$\eta$&$F(m_3)$&$F(m_2)$&$F(m_1)$\\
$\mcal L_{3,1}$&$F(m_2)\oplus F(m_3)$&$F(m_1)\oplus F(m_2)\oplus F(m_3)$&$F(m_1)\oplus F(m_2)$\\
$\mcal L_{5,1}$&$F(m_1)\oplus F(m_2)$&$F(m_1)\oplus F(m_2)\oplus F(m_3)$&$F(m_2)\oplus F(m_3)$
\end{tabular}.
\end{center}
\end{table}
\hspace{-17pt}Our routine exercise gives
\begin{equation}
    F(m)\cong1\oplus\mcal L_{5,1}\label{M38+12rank3alg1}
\end{equation}
with $m\cong m_1,m_3$ to realize $\text{FR}^{3,0}_1$. Similarly, the second solution gives
\begin{equation}
    F(m)\cong1\oplus\mcal L_{3,1}\label{M38+12rank3alg2}
\end{equation}
with $m\cong m_1,m_3$ to realize the same fusion ring. They both fail to be commutative.\footnote{They have double braidings
\[ c_{A,A}\cdot c_{A,A}\cong i\iota\cdot id_{3,1}\cdot p+(2-i)\iota\cdot id_{5,1}\cdot p, \]
and
\[ c_{A,A}\cdot c_{A,A}\cong(2+i)\iota\cdot id_{3,1}\cdot p-i\iota\cdot id_{5,1}\cdot p, \]
respectively.}

Finally, we consider the possibility of rank four. Fortunately, no computation is needed in this case; all candidate fusion rings require $\fp_{\mcal B}(A)=2+\sqrt2$, but (\ref{FPdimA}) and connectedness only allows candidates (\ref{M38+12rank3alg1},\ref{M38+12rank3alg2}). We get no new solution.\footnote{Of course, one can solve four-dimensional NIM-reps. Actually, we found many solutions. Interestingly, the solutions exist only when
\[ n_1=1_4=n_\eta. \]
This means $\mcal B$-symmetric gapped states described by rank four $\mcal B_A$ preserves the $\mbb Z_2$ symmetry, and its degenerated ground states is a consequence of spontaneously broken non-invertible symmetries $\mcal L_{3,1},\mcal L_{5,1}$. However, they require forms
\[ 1\oplus\eta\oplus a\mcal L_{3,1}\oplus a\mcal L_{5,1} \]
with $a\in\mbb N$. This cannot solve (\ref{FPdimA}) for $\text{FR}^{4,0}_1,\text{FR}^{4,2}_1$.}

To summarize, we found
\begin{table}[H]
\begin{center}
\begin{tabular}{c|c|c}
    Connected étale algebra $A$&$\mcal B_A$&$\rank(\mcal B_A)$\\\hline
    1&$\mcal B$&4
\end{tabular}.
\end{center}\caption{Results on $M(3,8)+\phi_{1,2}$}\label{M38+12results}
\end{table}
\hspace{-17pt}Note the absence of naive algebra $A\cong1\oplus\eta$. Since the $\mbb Z_2$ symmetry is anomaly-free, the algebra does exist. Our analysis suggests the $\mbb Z_2$ algebra would give $\mcal B_A$ with multiplicity.

Let us comment on physical implications of our results. Assuming the $\mcal B$-symmetric gapped phases are described by $\mcal B_A$'s, we get
\begin{equation}
    \text{GSD}=4.\label{GSDM38+12}
\end{equation}
This suggests $\mcal B$ is spontaneously broken in gapped phases.

\section{Discussion}
We developed a systematic method to classify connected étale algebras in pre-modular categories including degenerate non-pseudo-unitary ones. The assumption of spherical ambient category is solely motivated by physics examples, and mathematically it was unnecessary. Indeed, we did not use the assumption in our method. Therefore, it would be interesting to apply our method to non-spherical BFCs. In order to translate mathematical results to physical implications, we had to impose a few assumptions such as finiteness of module categories. It is desirable to justify this assumption (or clarify when this property holds).

Below, we list some future directions.
\begin{itemize}
    \item Generalized gauging: As we mentioned several times in the body, algebras can be interpreted as ways to gauge \cite{FRS02,CR12,BT17}. Thus, our procedure also provides sub-symmetries which can be gauged in a categorical symmetry. It would be interesting to find the full collection of algebras, and study gauged theories.
    \item Axiomatic massive RG flow: RG flows between RCFTs have been axiomatized as Kan extensions \cite{KKARG}. A natural question is this: mathematically, what is a massive RG flow?
    \item Higher dimensions: As we recalled in the introduction, our ultimate goal is to show IR behaviors of four-dimensional QCD. If we assume the IR theory is gapped, then suitable generalization of our method to four dimensions would bring us closer to the goal. The method proposed recently \cite{BBPS23,BBPS23'} may be useful.
\end{itemize}

We hope we could make progress in these questions in the future.

\section*{Acknowledgment}
We thank Victor Ostrik and Sebastien Palcoux for telling us important references. We also thank Victor Ostrik and Yuji Tachikawa for their comments on the draft.

\appendix
\setcounter{section}{0}
\renewcommand{\thesection}{\Alph{section}}
\setcounter{equation}{0}
\renewcommand{\theequation}{\Alph{section}.\arabic{equation}}

\section{Numerical check}\label{numericalcheck}
For unitary minimal models, it is not hard to check GSDs numerically. In this appendix, we perform the computation. We use the truncated conformal space approach (TCSA) \cite{YZ89}. For the details of our code based on \cite{STRIP}, see \cite{KKSUSY}.

The idea of the method is the following. We put our two-dimensional RCFTs on a cylinder with circumference $R$. A circle is viewed as a time slice, and we quantize our theory on it. We get the Hilbert space $\mcal H$. The full Hamiltonian $H$ of the deformed theory is given by
\begin{equation}
    H=H_\text{UV}-\sum_i\lambda_i\int_{\mbb S^1}\phi_i,\label{fullH}
\end{equation}
where $H_\text{UV}$ is the Hamiltonian of the UV RCFT and $\phi_i$'s are deformation operators with coupling constants $\lambda_i$'s. By diagonalizing the Hamiltonian, we obtain energies as functions of $R$ and we can extract GSDs from $R\to\infty$ limit corresponding to IR. However, this problem is intractable because the Hilbert space is infinite dimensional. Yurov and Zamolodchikov thus suggested to truncate the Hilbert space \cite{YZ89}. Then, it reduces to finite dimensional truncated Hilbert space $\mcal H_\text{truncated}$. Now, we can diagonalize the Hamiltonian acting on $\mcal H_\text{truncated}$ on a computer. Below, we show some lowest energy eigenvalues $E(R)$'s as functions of $R$. We read off GSDs from $R\to\infty$ limits.

\subsection{$M(5,4)+\phi_{1,3}$}
\begin{figure}[H]
\begin{center}
\includegraphics[width=0.8\textwidth]{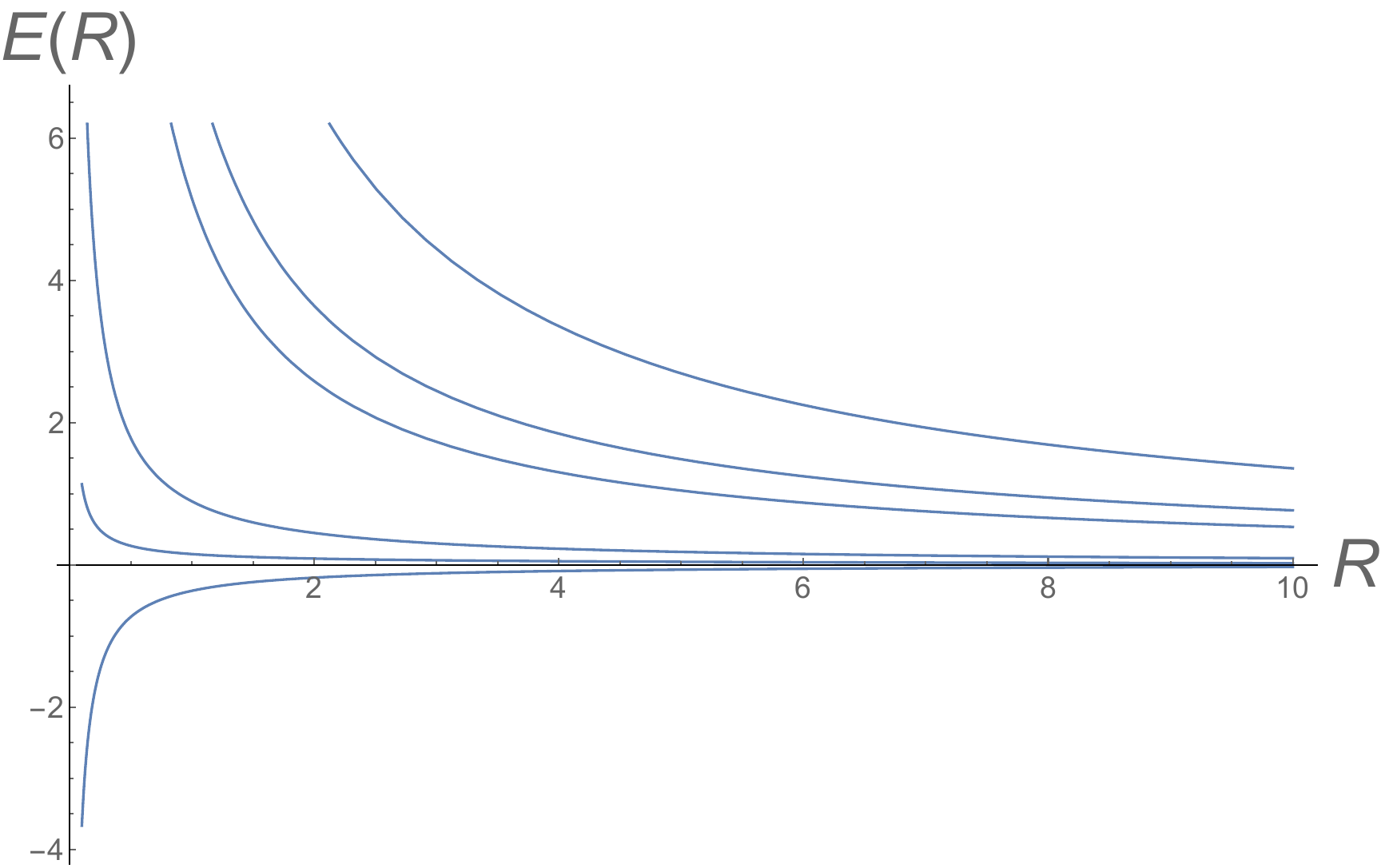}
    \caption{$\lambda_{1,3}<0$}\label{M54-13}
\end{center}
\end{figure}
We read off GSD from large $R$ region corresponding to IR. We see the three lowest energy eigenvalues approach each other. The numerical result suggests
\begin{equation}
    \text{GSD}=3.\label{M54+13TCSA}
\end{equation}

\subsection{$M(6,5)+\phi_{1,3}$}
\begin{figure}[H]
\begin{center}
\includegraphics[width=0.8\textwidth]{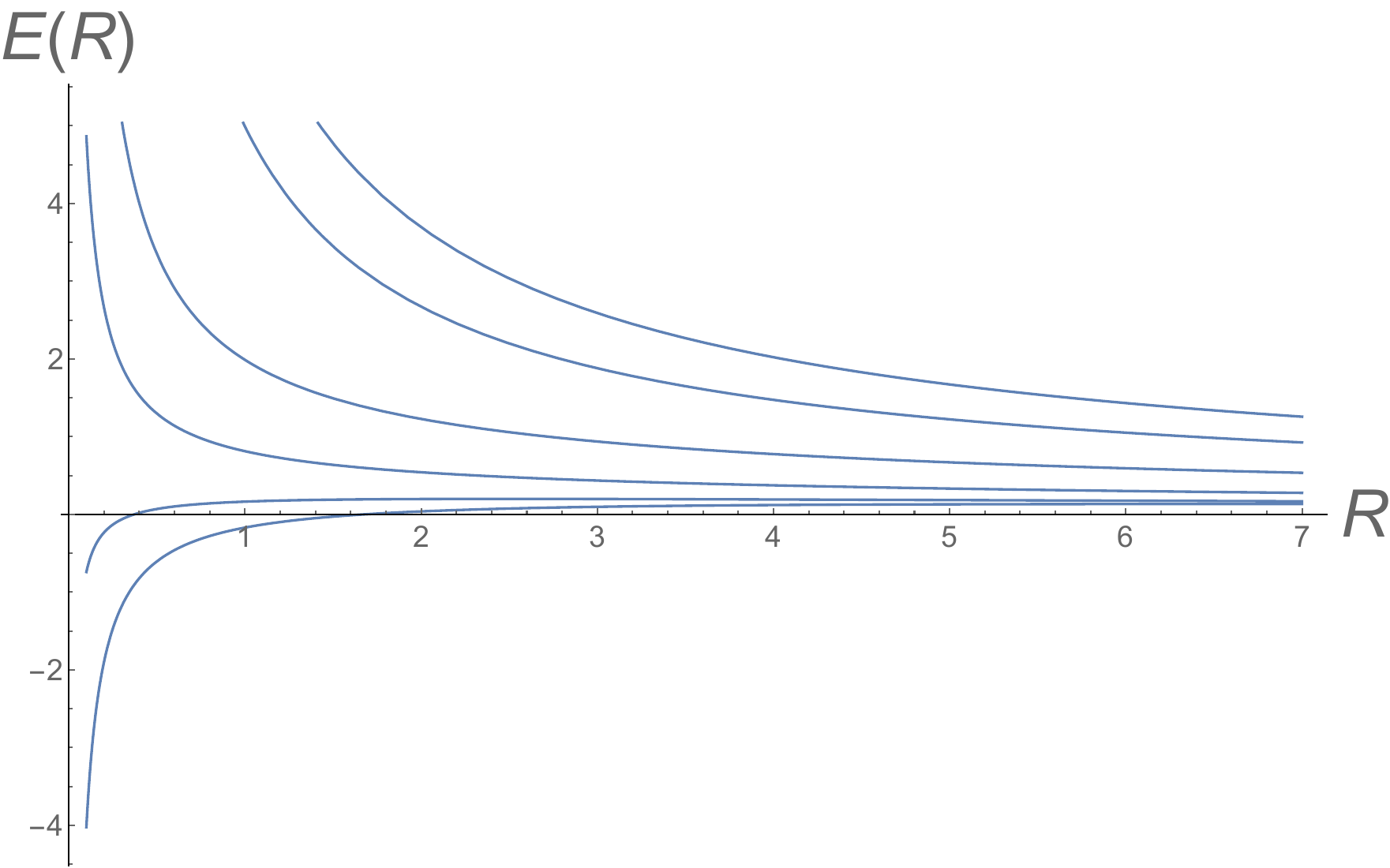}
    \caption{$\lambda_{1,3}<0$}\label{M65-13}
\end{center}
\end{figure}
The convergence is not as good as the other two examples. This would be because the conformal dimension $h_{1,3}=\frac23$ of the deformation operator is closer than the other examples to the threshold $3/4$ pointed out in \cite{GW11}. (Our code use their improvement via coupling constant renormalization.) Accordingly, a reader may think the third lowest energy eigenstate is also the ground state. However, odd GSDs are ruled out, and we conclude
\begin{equation}
    \text{GSD}=2.\label{M65+13TCSA}
\end{equation}

\subsection{$M(6,5)+\phi_{2,1}$}
\begin{figure}[H]
\begin{tabular}{cc}\hspace{-50pt}
    \begin{minipage}[t]{0.6\hsize}
    \centering
    \includegraphics[width=0.8\textwidth]{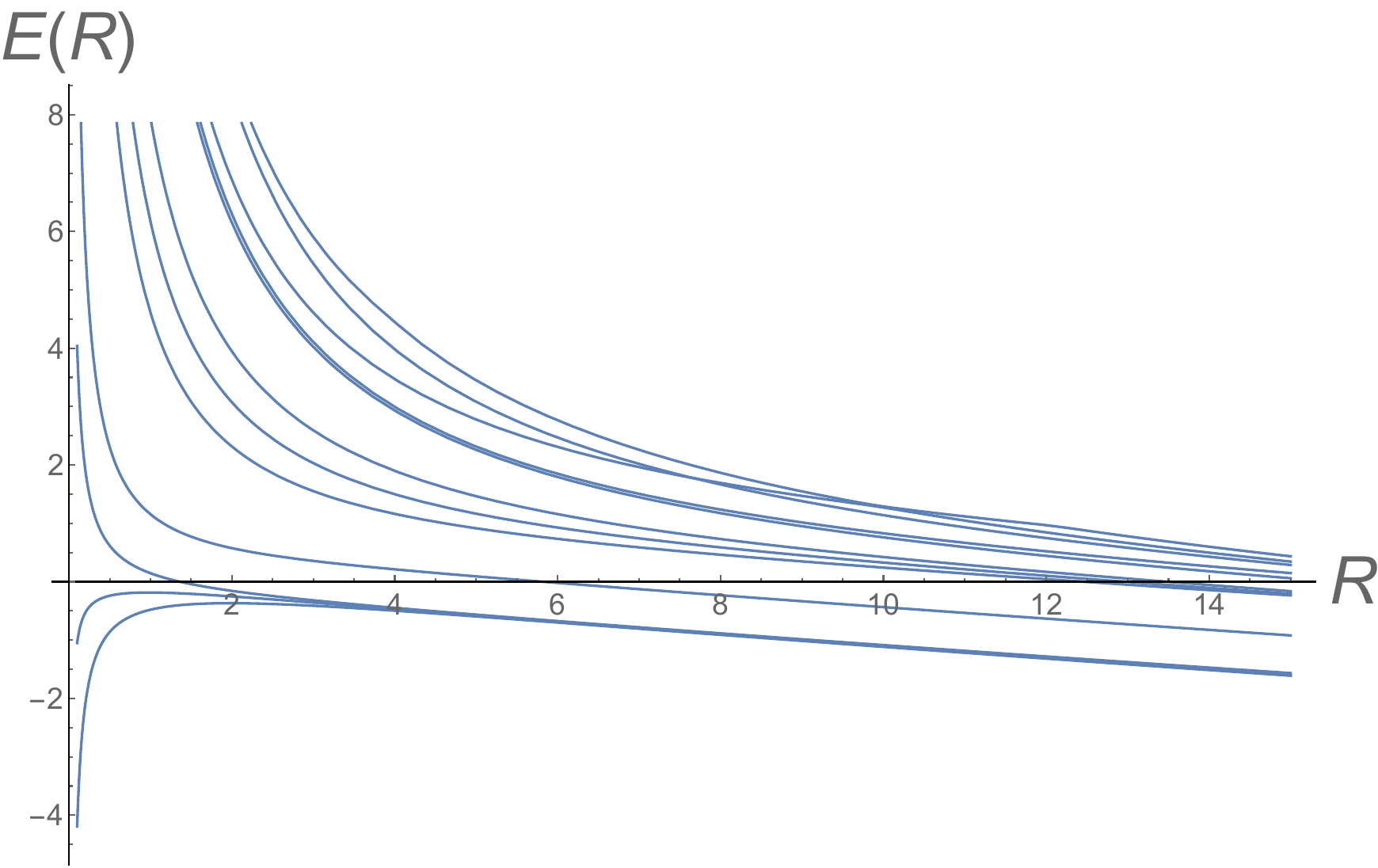}
    \subcaption{$\lambda_{2,1}>0$}\label{M65+21}
    \end{minipage}
    \begin{minipage}[t]{0.6\hsize}
    \centering
    \includegraphics[width=0.8\textwidth]{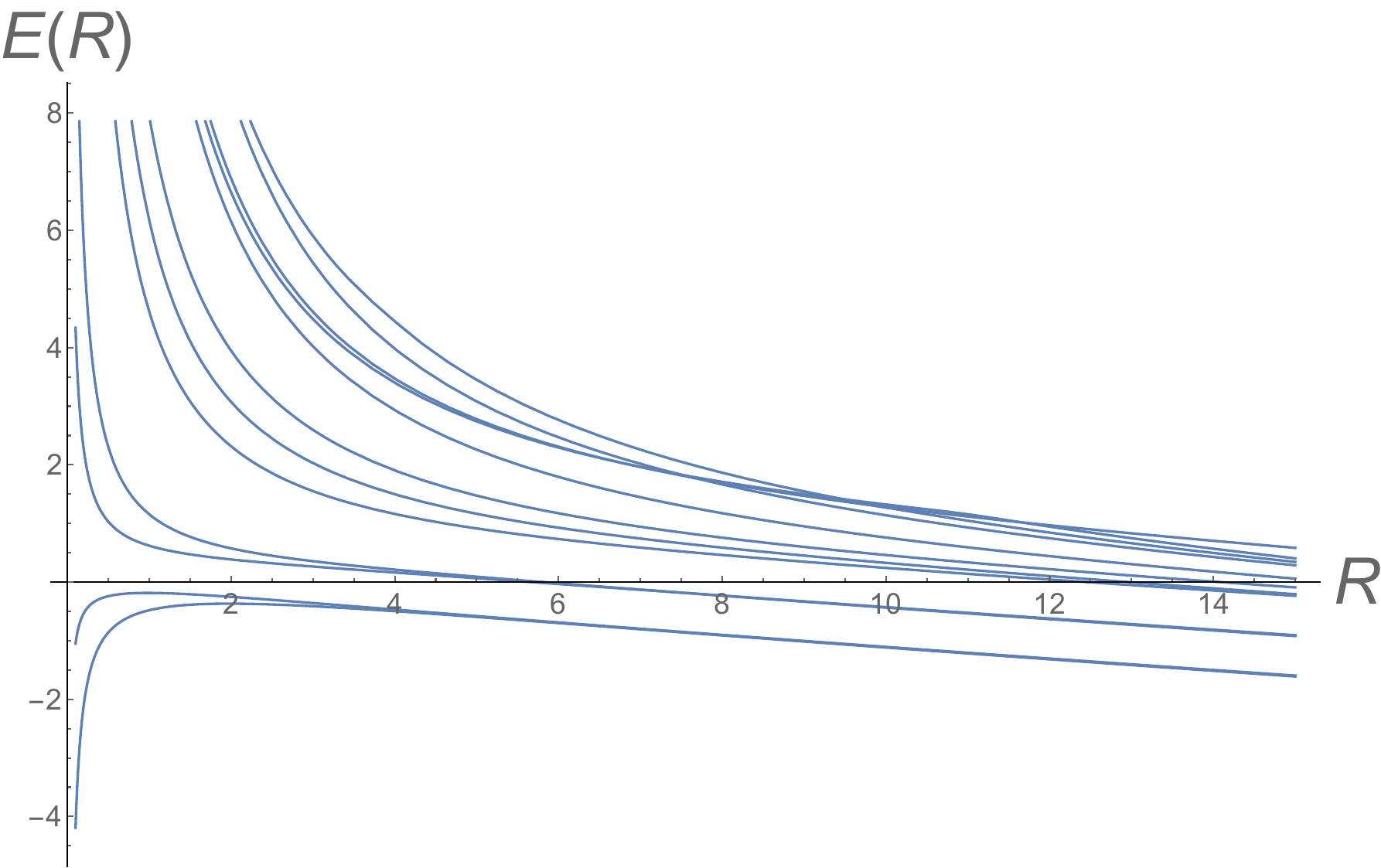}
    \subcaption{$\lambda_{2,1}<0$}\label{M65-21}
    \end{minipage}
\end{tabular}\caption{TCSA results}
\end{figure}
Since the conformal dimension $h_{2,1}=\frac25$ of the deformation operator is far from the threshold, we see beautiful convergence. The numerical results suggest
\begin{equation}
    \text{GSD}=\begin{cases}3&(\lambda_{2,1}>0),\\2&(\lambda_{2,1}<0).\end{cases}\label{M65+21TCSA}
\end{equation}

\end{document}